\documentclass[structabstract,onecolumn]{aa}
\usepackage{natbib}

\usepackage{hyperref}
\usepackage{amsmath}
\hypersetup{colorlinks=true,citecolor=blue}
\usepackage{draftcopy}

\usepackage{url}
\usepackage{amsmath}
\usepackage{amssymb}
\usepackage{amsfonts}
\usepackage[ruled]{algorithm2e}
\usepackage{graphicx,epsfig,graphics}
\usepackage{aas_macros}
\usepackage{algorithmic}
 
\newcommand{\norm}[1]{\left\lVert #1 \right\rVert}

\newcommand{\var}[1]{{\mathrm{Var}}\left[#1\right]}

\newcommand{\parenth}[1]{\left(#1\right)}
\newcommand{\crochets}[1]{\left[#1\right]}

\def\be{\begin{eqnarray}}
\def\ee{\end{eqnarray}}

\title{True CMB Power Spectrum Estimation}
\author{P. Paykari \inst{1} \thanks{paniez.paykari@cea.fr} \and  J.-L. Starck \inst{1} \and M. J. Fadili \inst{2}}
\institute{$^1$ Laboratoire AIM, UMR CEA-CNRS-Paris 7, Irfu, SAp/SEDI, Service d'Astrophysique, CEA Saclay, F-91191 GIF-SUR-YVETTE CEDEX, France. \\
$^2$ GREYC CNRS UMR 6072, ENSICAEN, 6 Bd du Mar\'echal Juin, 14050 Caen Cedex, France}

\begin{document}

\label{firstpage}
\date{\today}
%%% BEGIN DOCUMENT

%%%%%%%%%%%%%%%%%%%%%%%%%%%%%%%%%%%%%%%%%%%%%%%%%%%%%%
\abstract{{
} }
{
The cosmic microwave background (CMB) power spectrum
is a powerful cosmological probe as
it entails almost all the statistical information of the CMB 
perturbations. Having access to only one sky, the CMB power
spectrum measured by our experiments is only a realization of the true underlying
angular power spectrum. In this paper we aim to recover
the true underlying CMB power spectrum from the one realization that we have
without a need to know the cosmological parameters.  
}
{
The sparsity of the CMB power spectrum is first investigated in two dictionaries; Discrete Cosine
Transform (DCT) and Wavelet Transform (WT). 
The CMB power spectrum can be recovered with only a few percentage of
the coefficients in both of these dictionaries and hence is very compressible in these dictionaries. 
}
{
We study the performance of these dictionaries in smoothing a set of simulated power spectra. Based on this, 
we develop a technique that estimates the true underlying CMB power spectrum from data, i.e. without a 
need to know the cosmological parameters. 
}
{
This smooth estimated spectrum can be used to simulate CMB maps with 
similar properties to the true CMB simulations with the correct cosmological parameters. This allows 
us to make Monte Carlo simulations in a given project, without having to know the cosmological parameters.
The developed IDL code, {\bf TOUSI}, for Theoretical pOwer spectrUm using Sparse estImation, will be released with the next 
version of ISAP.
}

\maketitle

%\tableofcontents

%%%%%%%%%%%%%%%%%%%%%%%%%%%%%%%%%%%%%%%%%%%%%%%%%%%%%%
\section{Introduction}

Measurements of the CMB anisotropies are powerful cosmological probes.
In the currently favored cosmological model, with the nearly Gaussian-distributed
curvature perturbations, almost all the statistical information are
contained in the CMB angular power spectrum. The
observed quantity on the sky is generally the CMB temperature anisotropy
$\Theta(\vec{p})$ in direction $\vec{p}$, which is described as
$T(\vec{p})=T_{CMB}[1+\Theta(\vec{p})]$. This field is expanded on
the spherical harmonic functions as
\begin{align}
\Theta(\vec{p})=\sum_{\ell=0}^{+\infty}\sum_{m=-\ell}^{\ell} a[\ell,m]Y_{\ell m}(\vec{p})~, \\
\text{ where } a[\ell,m] = \int_{\mathbb{S}^2} \Theta(\vec{p}) Y^*_{\ell m}(\vec{p}) d\vec{p}~,
\end{align}
$\mathbb{S}^2 \subset \mathbb{R}^3$ is the unit sphere, $\ell$ is the multipole moment which is related to the angular
size on the sky as $\ell\sim180^{\circ}/\theta$ and $m$ is the phase
ranging from $-\ell$ to $\ell$. The $a[\ell,m]$ are the spherical
harmonic coefficients of the (noise-free) observed sky.
For a Gaussian random field, the mean and covariance are sufficient statistics, meaning that they carry all the
statistical information of the field. In case where the random field has zero mean, $\mathbb{E}(a_{00}) = 0$ and 
the expansion can be started at $\ell=2$, neglecting the dipole terms, i.e. $\ell=1$\footnote{The dipole anisotropy is dominated by the Earth's motion in space and it is hence ignored.}. For $\ell \geqslant 2$, the triangular
array $(a[\ell,m])_{\ell,m}$ represents zero-mean, complex-valued random coefficients, with variance
\begin{equation}
\label{eq:almvar}
\mathbb{E}(|a[\ell,m]|^2 )= C[\ell] > 0~, 
\end{equation} 
where $C[\ell]$ is the CMB angular power spectrum, which only depends on $\ell$ due the isotropy assumption. 
Therefore, from \eqref{eq:almvar}, an unbiased estimator of $C[\ell]$ is given by the empirical power spectrum
\begin{equation}
\label{eq:CMB_PS}
\widehat{C}[\ell] = \frac{1}{2\ell+1}\sum_{m}\left|a[\ell,m]\right|^{2}\;.
\end{equation} 
Furthermore, as the random field is stationary, the spherical harmonic coefficients are uncorrelated, 
\begin{equation}
\mathbb{E}(a[\ell,m]a^*[\ell^\prime,m^\prime]) = \delta_{\ell\ell^{\prime}}\delta_{mm^{\prime}}C[\ell] ~.
\end{equation}
Since they are Gaussian they are also independent. 
The angular power spectrum depends on the cosmological parameters through an angular transfer
function $T_\ell(k)$ as 
\begin{equation}
C[\ell]=4\pi\int\frac{dk}{k}\; T_\ell^{2}(k)P(k)\;,\label{eq:CMB}
\end{equation}
where $k$ defines the scale and $P(k)$ is the primordial matter power spectrum.

Making accurate measurements of this power spectrum has been one of the main
goals of cosmology in the past two decades. We have seen a range of
ground- and balloon-based experiments, such as Acbar \citep{ACBARl}
and CBI \citep{CBI-Readhead}, as well as satellite experiments, such
as WMAP \citep{WMAPBennett} and the recently launched satellite Planck
\citep{Planck}. All these experiments produce a temperature map of
the sky from which the CMB power spectrum is obtained. The estimation
of the power spectrum from CMB experiments is of great importance,
as this spectrum is a way to estimate the cosmological parameters
that describe the Universe. General methods for extracting this spectrum
from a $N_{pix}$-map, with nonuniform coverage and correlated noise
are quite expensive and time-consuming. Especially in the case of
Planck where we will be dealing with $N_{pix}=5\times10^{7}$. All
these experiments estimate the CMB angular power spectrum from a sky
map, which is a \emph{realization} of the underlying true power
spectrum; no matter how much the experiments improve, we are still
limited to an accuracy within the cosmic variance. This means that
even if we had a perfect experiment (i.e. with zero instrumental noise)
we would not be able to recover a perfect power spectrum due to the
cosmic variance limit. 

In this paper we investigate the possibility of estimating the true
underlying power spectrum from a \emph{realized} spectrum; 
an estimation of the true power spectrum without
a need to know the cosmological parameters. For this we exploit the
sparsity properties of the CMB power spectrum, and capitalize on it to propose an estimator
of the theoretical power spectrum. 

The idea of sparsity in different dictionaries has previously been used. 
For example, \citep{gauss:pia04} use wavelets to estimate the level of non-Gaussianity in the first year WMAP data. 
In \citep{astro:pia03} wavelets were used to estimate the primordial power spectrum. \citep{fay08}
have used wavelets to estimate the power spectrum from a CMB map, as an alternative to the MASTER method of \citep{master:hivon02}.
There has also been previous attempts to smooth the CMB power spectrum, for e.g. by the use of spline-fitting,
where the smoothed power spectrum has been used for visual aids \citep{Oh99_SmoothPS}! 
\\

In this paper the sparsity of the CMB power spectrum is used as a key ingredient in order to estimate
the theoretical power spectrum without having to know the cosmological parameters; 
this estimate will not belong to a set of possible theoretical power spectra (i.e. all $C[\ell]$  that
can be obtained by CAMB\footnote{CAMB solves the Boltzmann equations for a cosmological model set out by the given cosmological parameters.} by varying the cosmological parameters). Instead, 
such an estimation should be useful for other applications, such as:
\begin{itemize}
\item[$\bullet$] Monte Carlo: we may want to make Monte Carlo simulations in some applications without 
assuming the cosmological parameters.
\item[$\bullet$] Wiener filtering: Wiener filtering is often used to filter the CMB map and it requires the theoretical 
power spectrum as an input. We may not want to assume any cosmology at this stage of the processing.
\item[$\bullet$] Some estimators (weak lensing, ISW, etc.) require the theoretical power spectrum to be known. Using a
data-based estimation of the theoretical $C[\ell]$ could
be an interesting alternative, or at least a good first guess in an iterative scheme where the theoretical $C[\ell]$ is required 
to determine the cosmological parameters.
\end{itemize}

\subsection*{Paper content}
Section~\ref{sect_sparsity} introduces the concepts of sparse representation and its applications to the CMB 
power spectrum. In Section~\ref{sec:estimator} we explain how sparsity is used to propose an estimator of 
the theoretical power spectrum. Experimental results are described and discussed in Section~\ref{sec:results}, 
and Section~\ref{sec:discussion} is devoted to a discussion and comparison to the moving average estimator. 
In Section~\ref{sec:conclusion}, conclusions are drawn and potential perspectives are stated.

%%%%%%%%%%%%%%%%%%%%%%%%%%%%%%%%%%%%%%%%%%%%%%%%%%%%%%
\section{Sparsity of the CMB Power Spectrum}
\label{sect_sparsity}

\subsection{A brief tour of sparsity}
A signal $X=(X[1],\ldots,X[N])$ considered as a vector in $\mathbb{R}^N$, is said to be sparse if
most of its entries are equal to zero. If $k$ number
of the $N$ samples are not equal to zero, where $k\ll N$, then the signal
is said to be $k$-sparse. In the case where only a few of the entries
have large values and the rest are zero or close to zero the signal
is said to be weakly sparse (or compressible). With a slight abuse of terminology, in the sequel, 
we will call compressible signals sparse.
Generally signals are not sparse in direct space, but can be sparsified by transforming
them to another domain. For example, $\sin(x)$ is $1$-sparse
in the Fourier domain, while it is clearly not sparse in the original
one. In the so-called sparsity synthesis model, a signal can be represented as the linear expansion 
\begin{equation}
X =\Phi\alpha=\sum_{i=1}^{T}\phi_{i}\alpha[i]\ ~,
\end{equation}
where $\alpha[i]$ are the synthesis coefficients of $X$, $\Phi=(\phi_{1},\ldots,\phi_{T})$ is the dictionary, 
and  $\phi_{i}$ are called the atoms (elementary waveforms) of the dictionary $\Phi$.
In the language of linear algebra, the dictionary $\Phi$ is a $N\times T$ matrix whose columns are
the atoms normalized, supposed here to be normalized to a unit $\ell_{2}$-norm, i.e. 
$\forall i\in[1,T],\left\Vert \phi_{i}\right\Vert _2^2=\sum_{n=1}^{N}\left|\phi_{i}[n]\right|^{2}=1$\footnote{The $l_p$-norm of a vector $X$, $p \geq 1$, is defined as $\left\Vert X \right\Vert _p=\left(\sum_i \left| X[i] \right|^p\right)^{1/p}$, 
with the usual adaptation $\|X\|_\infty=\max_i X[i]$.}.
A function can be decomposed in many dictionaries, but the best dictionary is the one with the 
sparsest (most economical) representation of the signal. 
In practice, it is convenient to use dictionaries with fast implicit transform (such as Fourier transform, wavelet transform, etc.) 
which allow us to directly obtain the coefficients and reconstruct the signal from these coefficients using fast algorithms 
running in linear or almost linear time (unlike matrix-vector multiplications).
The Fourier, wavelet and discrete cosine transforms provide certainly the most well known dictionaries. 
A comprehensive account on sparsity and its applications can be found in the monograph \citep{starck:book10}.

%%%%%%%%%%%%%%%%%%%%%%%%%%%%%%%%%%%%%%%%%%%%%%%%%%%%%%
\subsection{Which Dictionary for the Theoretical CMB Power Spectrum?}
\label{which_dico}
We investigate the sparsity of the CMB power spectrum in two different
dictionaries, both having a fast implicit transform: 
the Wavelet Transform (WT) and the Discrete Cosine Transform (DCT). 

\begin{figure*}[htb]
\centering{
\includegraphics[width=\linewidth]{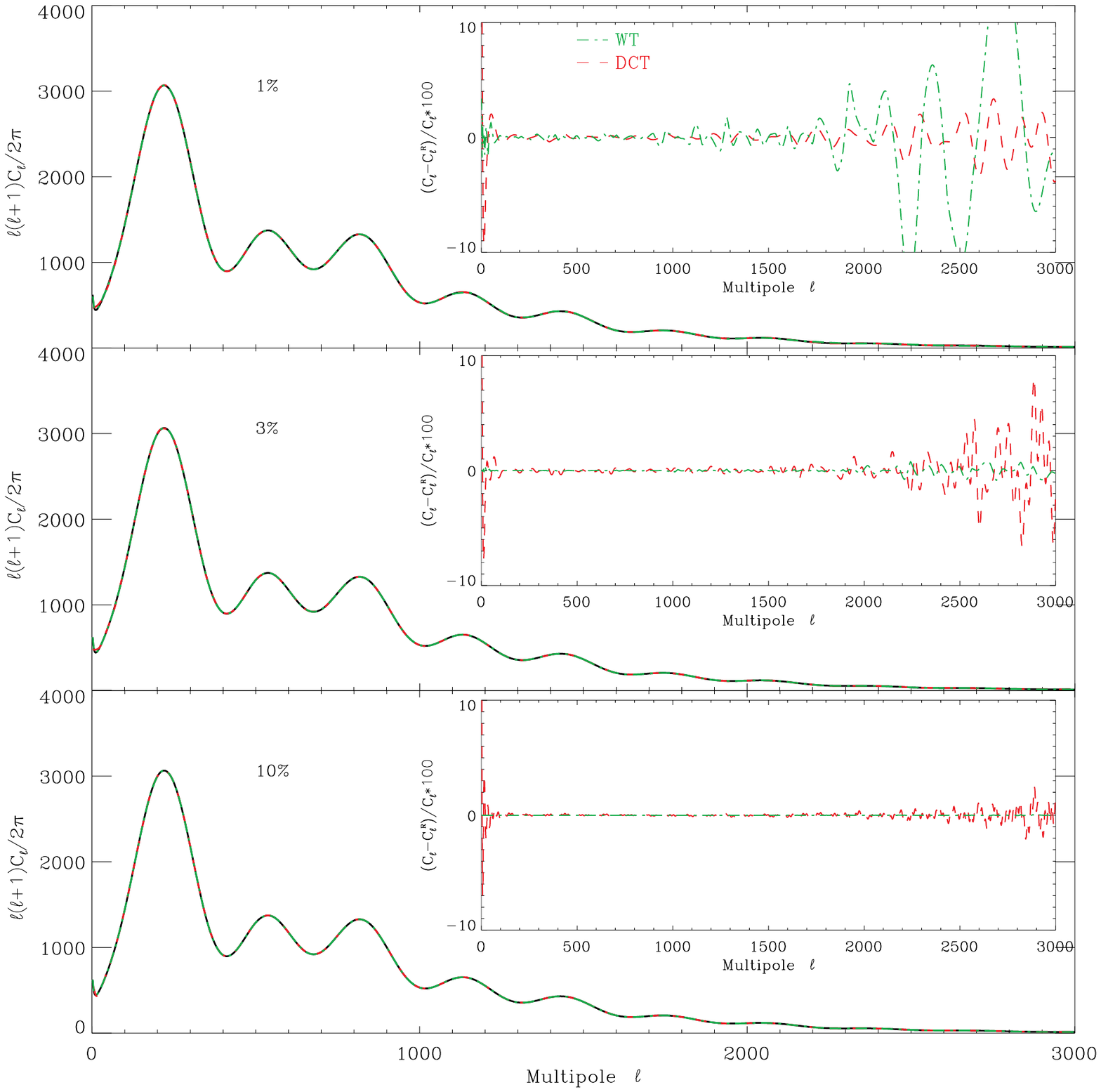}
\caption{A theoretical CMB power spectrum along with
the reconstructed power spectra, using the DCT and WT dictionaries. The
panels show the reconstructions for different fractions of the coefficients
used. The inner plots show the differences between the actual and
the reconstructed power spectra. Both dictionaries suffer from boundary effects, but this is 
more severe for DCT as the corresponding atoms are not compactly supported. 
It is worth mentioning that the power spectrum
that is decomposed onto the two dictionaries is in the form $\ell(\ell+1)C[\ell]/2\pi$.}
\label{Fig1}}
\end{figure*}

\begin{figure*}[htb]
\centering{
\includegraphics[width=\linewidth]{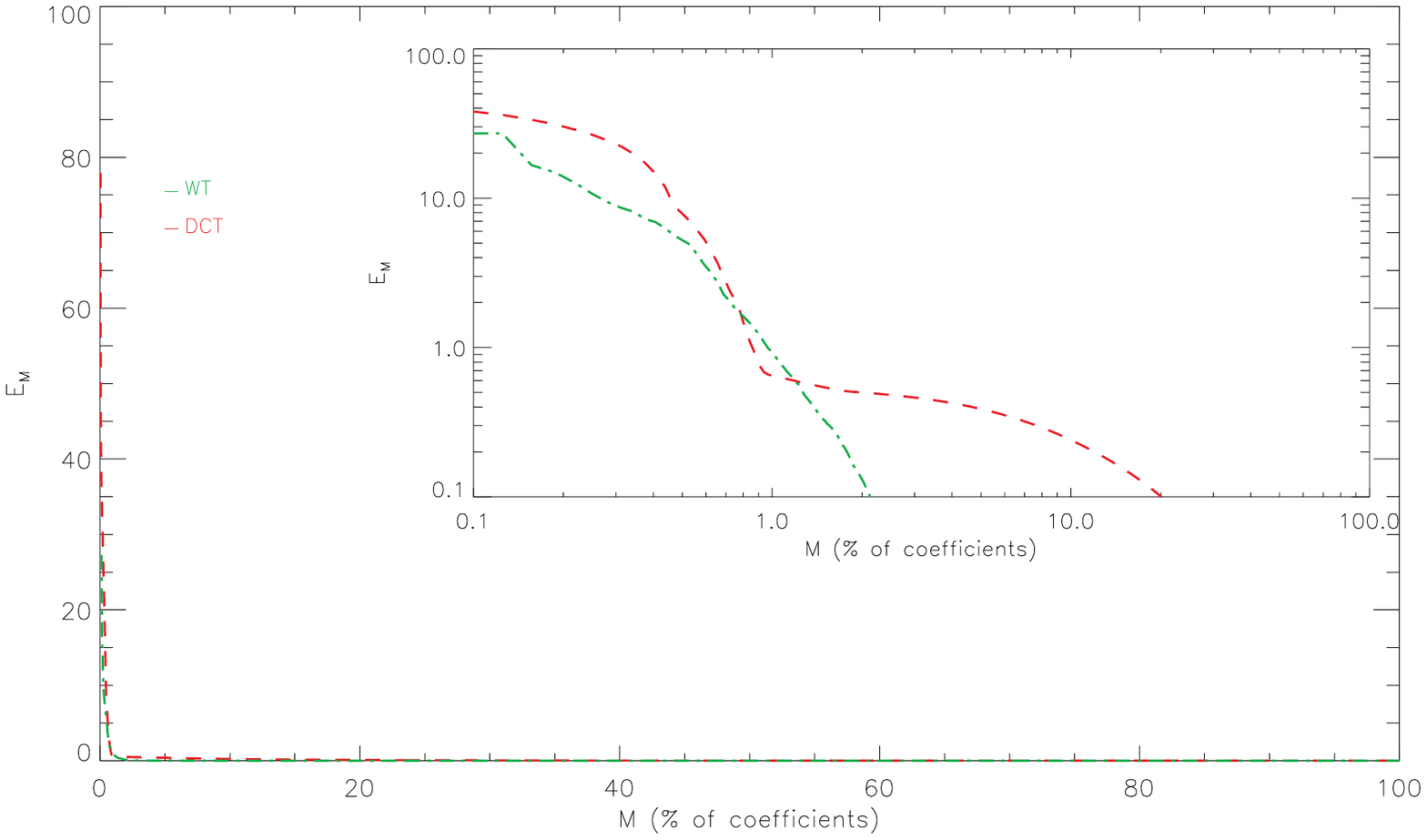}
\caption{Non-Linear Approximation (NLA) error curves for the two dictionaries.
Below $1\%$ the DCT curve is dropping faster, which means it is doing
a better job. However, past $\sim2\%$ the DCT curve flattens off
while WT decreases to $\sim0$ very quickly.}
\label{Fig2}}
\end{figure*}

Figure~\ref{Fig1} shows an angular power spectrum
(calculated by CAMB \citep{camb} with WMAP$7$ \citep{WMAP7} parameters)
along with the DCT- and WT-reconstructed power spectra with a varying fraction of
the largest transform coefficients retained in the reconstruction. The
inner plots show the difference between the actual power spectrum
and the reconstructed ones. It can be seen that with only a few percentage
of the coefficients the shape of the power spectrum is correctly
reconstructed in both dictionaries. The height and the position of the peaks and troughs are
of great importance here as the estimation of the cosmological parameters
heavily relies on these characteristics of the power spectrum. The
best domain would be the one with the sparsest representation and
yet the most accurate representation of
the power spectrum. Let ${{C}[\ell]}^{(M)}$ be its best $M$-term approximation, i.e. obtained by reconstructing 
from the $M$-largest (in magnitude) coefficients of ${C}[\ell]$ in a given domain. To compare the WT and DCT 
dictionaries, we plot the resulting non-linear approximation (NLA) error curve in Figure~\ref{Fig2}, which shows 
the reconstruction error $E_M$ as a function of $M$, the number of retained coefficients; 
\begin{equation}
E_{M}= {\frac{\left\Vert {C}[\ell]-{{C}[\ell]}^{(M)}\right\Vert _2}{\left\Vert C[\ell]\right\Vert _2}}\times100\;.
\label{NLA}
\end{equation}
As $M$ increases we get closer to the complete reconstruction and
 the error reaches $0$ when all the coefficients have been used. Usually the domain
with the steepest $E_M$ curve is the sparsest domain. In this case though both dictionaries have very similar behaviors. 
There is only a small window in the coefficients for which DCT does a better
job than WT. However, DCT flattens after using $\sim1\%$ of the coefficients 
and does not improve the reconstruction until a big proportion of the
coefficients have been used. 

Both dictionaries seem to suffer from boundary issues at low and high $\ell$s.
This can be solved for high $\ell$s as one can always perform the reconstruction
beyond the desired $\ell$. For low $\ell$s it can be solved by different means, 
such as extrapolation of the spectrum. Note that the boundary issues are more severe in
the DCT domain than WT; this is due to the fact that DCT atoms are not compactly supported.

\begin{figure*}[htb]
\centering{
\includegraphics[width=\linewidth]{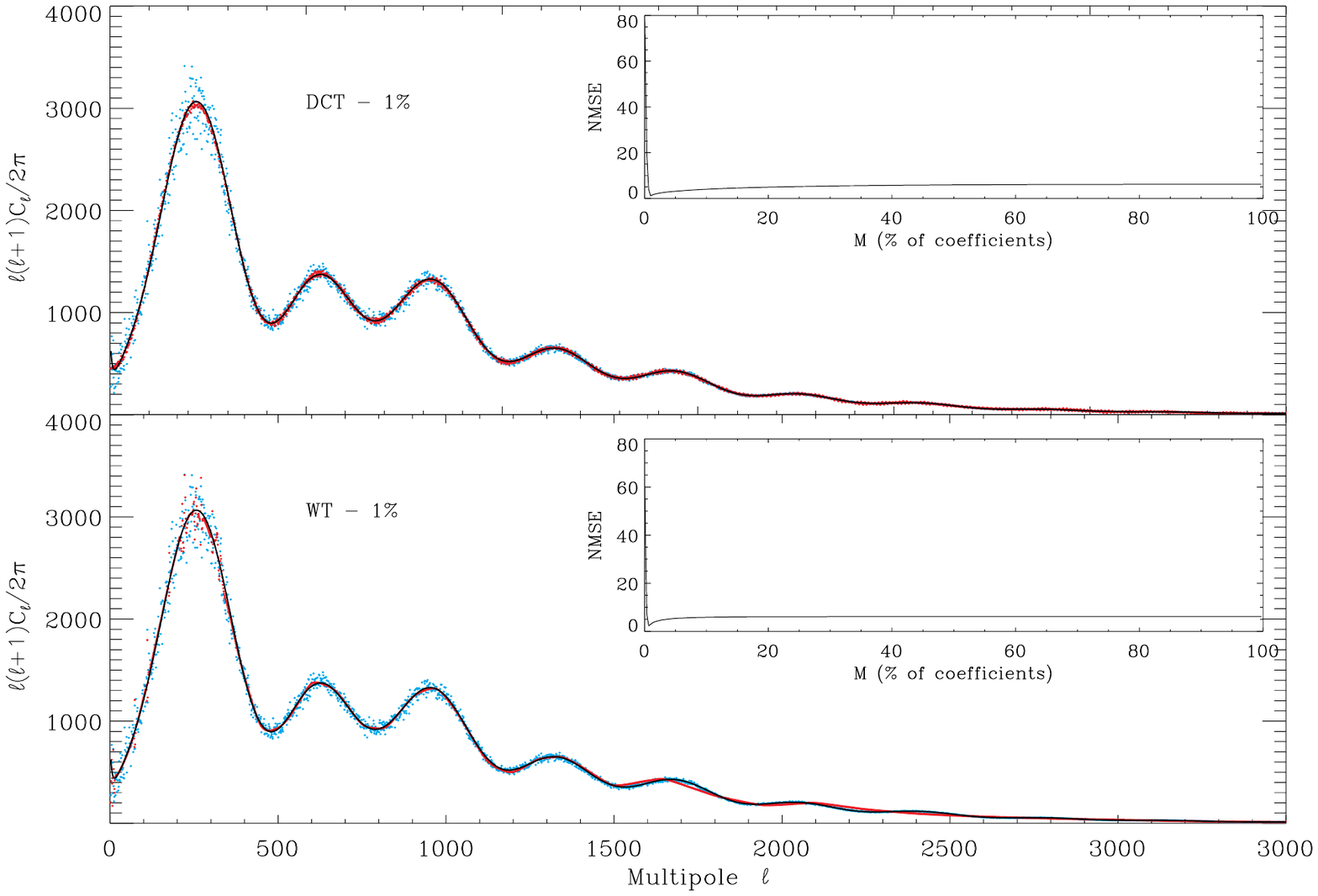}
\caption{A simulated CMB power
spectrum along with the reconstructed spectra, using the DCT and WT dictionaries. 
The black solid line is the true underlying power spectrum
from which the simulations were made. The blue and red dots show the simulated
and the reconstructed power spectra respectively. With only $1\%$
of the coefficients, DCT can recover the input power spectrum (i.e.
the black solid line) very well, recovering the peaks and troughs
accurately. Unlike DCT, WT seems to have difficulties
in recovering the peaks and troughs. The inner plots shows the NMSE curves.}
%\label{Fig3WTDCT}
\label{Fig3}
}
\end{figure*}

\begin{figure*}[htb]
\centering{
\includegraphics[width=\linewidth]{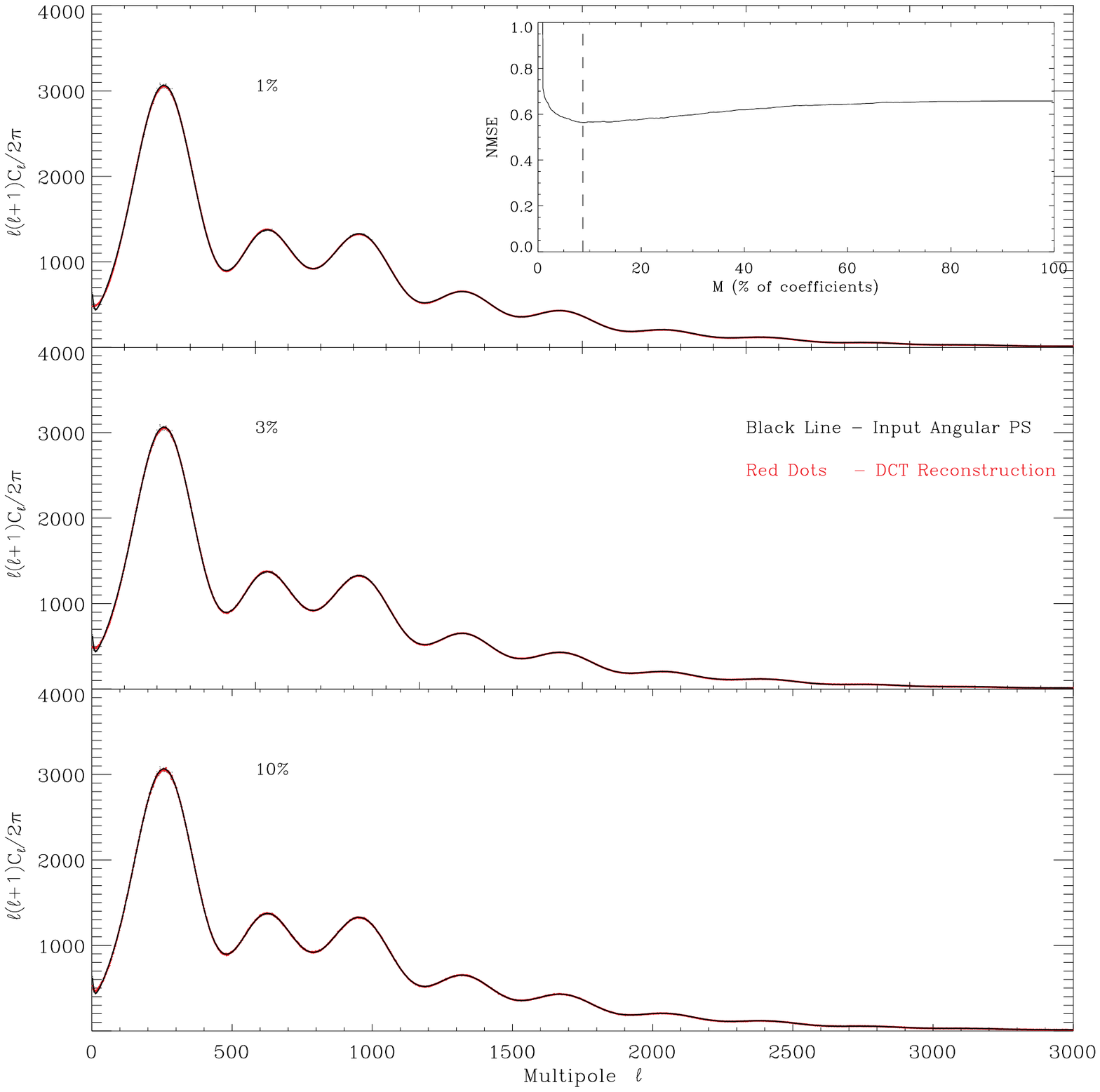}
\caption{The average of the DCT-reconstructed power spectra when different fraction
of most significant coefficients have been used in the reconstruction. The inner plot shows the NMSE curve for this average. 
The minimum of the curve is at less than $10\%$ of the coefficients, meaning that the true CMB spectrum 
can be recovered with less than $10\%$ of the coefficients while ensuring a good bias-variance tradeoff.}
%\label{Fig5DCT}
\label{Fig4}
}
\end{figure*}

\begin{figure*}[htb]
\centering{
\includegraphics[width=\linewidth]{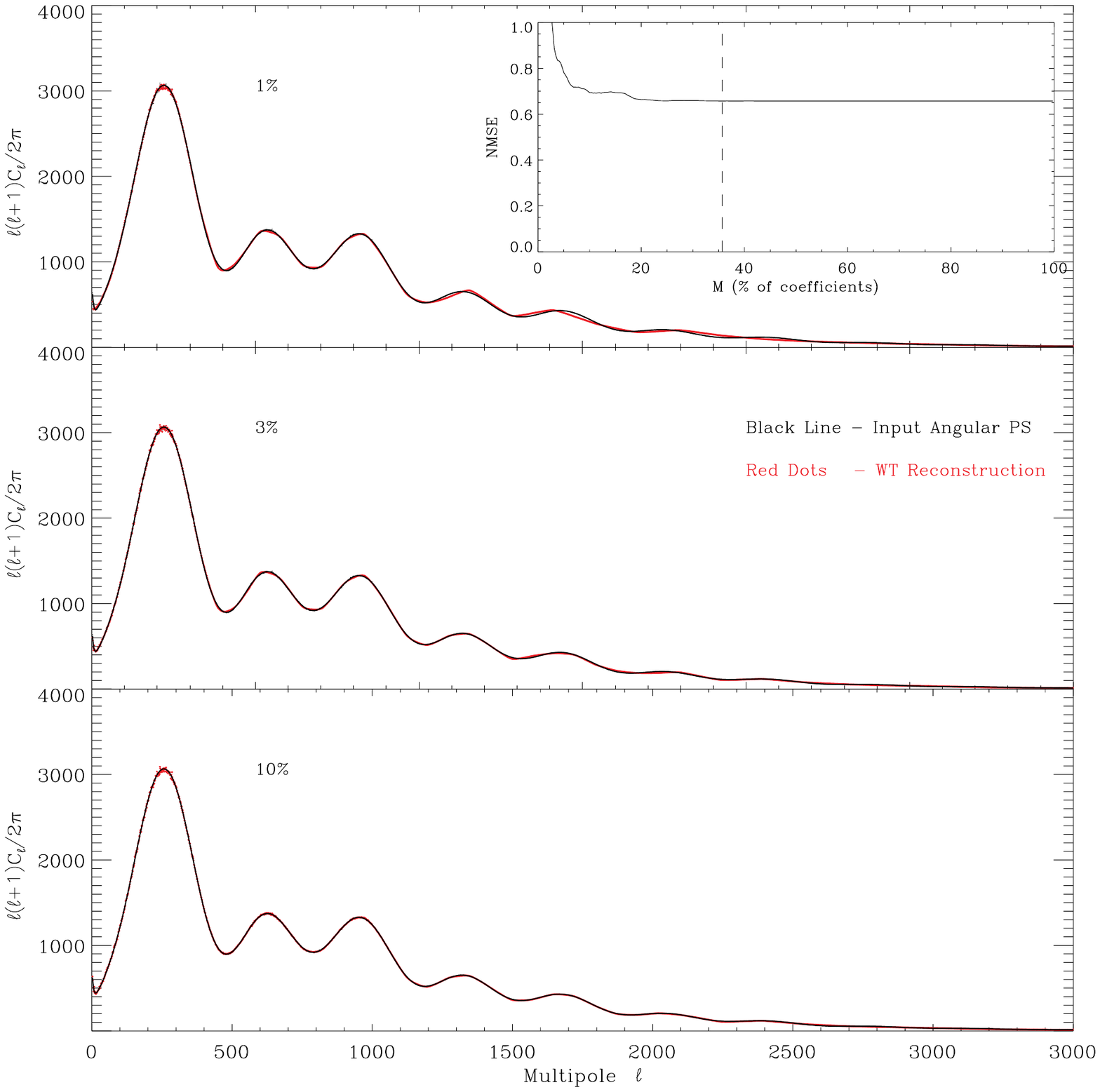}
\caption{Same as Figure~\ref{Fig4},
but for the WT domain. It seems that WT cannot do a great job for simulated power spectra, compared to the DCT domain; the minimum 
of the NMSE curve is at more than $35\%$ --- while the curve is pretty much flat after $\sim20\%$ anyway.}
%\label{Fig6WT}
\label{Fig5}
}
\end{figure*}

Next we investigate the sparsity of a set of \emph{realized} spectra in the two dictionaries. 
We simulate $100$ maps from the theoretical power spectrum used previously and estimate their power spectra
using equation \ref{eq:CMB_PS} .
As before, we decompose each realization in the DCT and WT dictionaries and reconstruct keeping increasing 
fractions of the largest coefficients. At this stage, it is important to note that, as we are dealing with the 
empirical power spectrum, we are no longer in an approximation setting but rather in an estimation one. 
Indeed, the empirical power spectrum can be seen as a noisy version of the true one. Intuitively, 
reconstructing from a very small fraction of high coefficients will reject most of the noise (low estimator variance) 
but at the price of retaining only a small fraction of the true spectrum coefficients (large bias). The converse 
is true when a large proportion of coefficients is kept in the reconstruction. Therefore, there will exist a threshold 
value that will entail a bias-variance tradeoff, hence minimizing the estimation risk. 
This is exactly the idea underlying thresholding estimators in sparsifying domains. 

This discussion is clearly illustrated by the inner plots of Figure~\ref{Fig3}, which shows the normalized mean-square error (NMSE) defined as 
\begin{equation}
\mathrm{NMSE}_{M} = {\frac{\left\Vert {C}[\ell]-{\widehat{C}[\ell]}^{(M)}\right\Vert _2}{\left\Vert C[\ell]\right\Vert _2}}\times100\;,
\end{equation}
as a function of the fraction of coefficients used in the reconstruction. The error is large when only a few coefficients are used. 
As more coefficients are included,
one starts to recover the main (i.e. the general shape of the spectrum)
features of the power spectrum. With more coefficients, more noise enters the estimation and the error increases again. 
The NMSE curve shows a clear minimum at which the underlying true power spectrum is best recovered. 

Despite the differences in the performance of the two dictionaries, the minima 
of the NMSE are around the same proportions of the coefficients. This is because 
the NMSE reflects a \emph{global} behavior. On the one hand, although the DCT can recover the features of the 
spectrum correctly, it is less smooth than WT. Conversely, the WT cannot reconstruct the proper shape of the power spectrum, but provides a smoother estimate. 

Figures \ref{Fig4} and \ref{Fig5} show the same results for an average over the $100$ realizations.
It can be seen that on average DCT does a great job in the 
recovery of the features of the CMB spectrum. Indeed the minimum of NMSE curve
is at a lower proportion of coefficients for DCT than WT. This is because the small \emph{noisy} features 
of the DCT-reconstructed spectra cancels out in the averaging, while the wrong recovery of the shape of the spectrum by WT does not. 
In a nutshell, DCT seems to do a better job in reconstructing the true underlying CMB power spectrum from its realizations. 

To summarize, from the above discussion, we conclude the following:
\begin{itemize}
\item[$\bullet$] the CMB power spectrum is very sparse in both the DCT and WT dictionaries, 
although their sparsifying capabilities are different; 
\item[$\bullet$] DCT recovers global features of spectrum (i.e. the peaks and troughs) while 
WT recovers localized features;
\item[$\bullet$] in the case of realizations, WT recovers more localized (noisy) features than 
the global ones, while the DCT concentrates on the global features. 
\end{itemize}

In the next section, these complementary capabilities of the DCT and WT transforms will be combined to 
propose a versatile way for adaptively estimating the theoretical power spectrum from a single realization of it. 

%%%%%%%%%%%%%%%%%%%%%%%%%%%%%%%%%%%%%%%%%%%%%%%%%%%%%%
\section{Sparse Reconstruction of the Theoretical Power Spectrum}
\label{sec:estimator}
Let's start with the simple model where the observed signal $Y$ is contaminated by a zero-mean white Gaussian noise, $Y = X + \varepsilon$, where 
$X$ is the signal of interest and $\varepsilon \sim \mathcal{N}(0,\sigma^2)$.
Sparse recovery with an analysis-type sparsity prior amounts to finding the solution of the following problem:
\begin{equation}
\label{eq_minl1_den}
\min_{X}  \norm{ \Phi^{T} X}_1\ ~~ s.t. ~~  \norm{Y - X}_2 \le \delta~, 
\end{equation}
where $\Phi^{T} X$ represent the transform coefficients of $X$ in the dictionary $\Phi$, and $\delta$ 
controls the fidelity to the data and obviously depends on the noise standard deviation $\sigma$.

Let's now turn to denoising the power spectrum from one empirical realization of it. In this case, however, 
the noise is highly non-Gaussian and needs to be treated differently. Indeed, as we will see in the next 
section, the empirical power spectrum will entail a multiplicative $\chi^2$-distributed noise with a number 
of degrees of freedom that depends on $\ell$. That is, the noise has a variance profile that dependents both on the true spectrum and $\ell$.
We therefore need to \textit{stabilize} the noise on the empirical power spectrum prior to estimation, 
using a Variance Stabilization Transform (VST). Hopefully, the latter will yield stabilized samples that 
have (asymptotically) constant variance, say $1$, irrespective of the value of the input noise level.

%%%%%%%%%%%%%%%%%%%%%%%%%%%%%%%%%%%%%%%%%%%%%%%%%%%%%%
\subsection{Variance Stabilizing Transform} \label{sect_vst}
In the statistical literature the problem of removing the noise from an empirical power spectrum goes by the 
name of periodogram denoising \citep{rest:donoho93_2}. In \citep{komm1999}, approximating the noise 
with a correlated Gaussian noise model, a threshold was derived at each wavelet scale using the MAD 
(Median of Absolute Deviation) estimator. A more 
elegant approach was proposed in \citep{rest:donoho93_2,rest:moulin94}, where the so-called Wahba VST was used. This VST is defined as:
\begin{equation}
\label{eq:wahba}
{\cal T}(X) = \left( \log X + \gamma \right) \frac{\sqrt6}{\pi}~,
\end{equation}
where $\gamma = 0.57721...$ is the Euler-Mascheroni constant. After the VST, the stabilized samples can be 
treated as if the noise contaminating them were white Gaussian noise with unit variance.

We will take a similar path here, generalizing the above approach to the case of the angular power spectrum. 
Indeed, from  \eqref{eq:CMB_PS}, one can show that under mild regularity assumptions on the true power spectrum,
\begin{equation}
\label{eq:Clasymp}
\widehat{C}[\ell] \overset{d}{\rightarrow} C[\ell] Z[\ell], \text{ where } \; \forall \ell \geq 2, ~ 2L Z[\ell] \sim \chi^2_{2L}, L=2\ell+1~.
\end{equation}
$\overset{d}{\rightarrow}$ means convergence in distribution. From \eqref{eq:Clasymp}, it is appealing then 
to take the logarithm so as to transform the multiplicative noise $Z$ into an additive one. The resulting $\log$-stabilized empirical power spectrum reads
\begin{equation}
\label{eq:vst}
C^{s}[\ell]  :=  {\cal T}_\ell (\widehat{C}[\ell]) = \log \widehat{C}[\ell] -\mu_L = \log C[\ell] + \eta[\ell] ~. 
\end{equation}
where $\eta[\ell] := \log Z[\ell] - \mu_L$, $L=2\ell+1$. Using the asymptotic results from \citep{bartlett46} on 
the moments of $\log-\chi^2$ variables, it can be shown that $\mu_L = \psi_0\parenth{L} - \log L$, $\mathbb{E}(\eta[\ell]) = 0$ and $\sigma_L^2=\var{\eta[\ell]} = \psi_1\parenth{L}$, where 
$\psi_m(t)$ is the standard polygamma function, $\psi_m(t)=\frac{d^{m+1}}{d t^{m+1}}\log \Gamma(t)$.

We can now consider the stabilized $C^{s}[\ell]$ as noisy versions of the $\log C[\ell]$, where the noise is 
zero-mean additive and independent. Owing to the Central Limit Theorem, the noise tends to Gaussian 
with variance $\sigma_L^2$ as $\ell$ increases. At low $\ell$, normality is only an approximation. In fact, 
it can be show that the noise $\eta[\ell]$ has a probability density function of the form
\begin{equation}
p_{\eta}(\ell) = \frac{(2L)^{L}}{2^L \Gamma(L)}\exp\crochets{L\parenth{\ell +\mu_L-e^{\ell + \mu_L}}}~, 
\end{equation}
which might be used to estimate the thresholds in the wavelet domain. 

In order to standardize the noise, the VST \eqref{eq:vst} will be slightly modified to the normalized form
\begin{equation}
\label{eq_logvar}
C^{s}[\ell]  :=  {\cal T}_\ell ( \widehat{C}[\ell]  )  = \frac{ \log \widehat{C}[\ell]-\mu_L}{\sigma_L} = X^{s}[\ell] + \varepsilon[\ell] \;.
\end{equation}
where now the noise $\varepsilon[\ell]$ is zero-mean (asymptotically) Gaussian with unit variance, and 
$X^{s}[\ell]:=\log C[\ell]/\sigma_L$. It can be checked that the Wahba VST \eqref{eq:wahba} is a specialization of \eqref{eq_logvar} to $L=0$.

In the following, we will use the operator notation ${\cal T}(X)$ for the VST that applies \eqref{eq_logvar} 
entry-wise to each $X[\ell]$, and ${\cal R}(X)$ its inverse operator, i.e. ${\cal R}(X):=\big({\cal R}_\ell (X[\ell])\big)_\ell$ with ${\cal R}_\ell (X[\ell]) = \exp ({\sigma_L}  X[\ell] )$.

%%%%%%%%%%%%%%%%%%%%%%%%%%%%%%%%%%%%%%%%%%%%%%%%%%%%%%
\subsection{Signal detection in the wavelet domain} \label{sect_noise}
Without of loss of generality, we restrict our description here to the wavelet transform. The same 
approach applies to other sparsifying transforms, e.g. DCT, just as well. 

In order to estimate the true CMB power spectrum from the wavelet transform, 
it is important to detect the wavelet coefficients which are ``significant'', i.e.   
the wavelet coefficients which have an absolute value too large to be 
due to noise (cosmic variance + instrumental noise). Let $w_j[\ell]$ the wavelet coefficient of a signal $Y$ 
at scale $j$ and location $\ell$. We define the multiresolution support $M$ of $Y$ as:
\begin{equation}
{M}_j[\ell] = 
  \begin{cases}
  1 & \mbox{ if }   w_{j}[\ell]  \mbox{ is significant,} \\
  0 & \mbox{ if }   w_{j}[\ell]  \mbox{ otherwise.}
  \end{cases}
\end{equation}
For Gaussian noise, it is easy to derive an estimation
of the noise standard deviation $\sigma_j$ at scale $j$ from the 
noise standard deviation, which can be evaluated with good accuracy in
an automated way \citep{starck:sta98_3}.
To detect the significant wavelet coefficients, 
it suffices to compare the wavelet coefficients in magnitude $|w_{j}[\ell]|$ to a threshold level
$t_j$. This threshold is generally taken to be equal to $\kappa \sigma_j$, where 
$\kappa$ ranges from 3 to 5. 
This means that a small magnitude compared to the threshold implies that the coefficients is 
very likely to be due to noise and hence insignificant. Such a decision rule corresponds to the hard-thresholding operator
\begin{eqnarray}
\begin{array}{l}
\mbox{ if }  \left| w_{j}[\ell]  \right| \ \geq \  t_j  \ \ \mbox{ then } w_{j}[\ell]  \mbox{ is significant }, \\ 
\mbox{ if }  \left| w_{j}[\ell]  \right| \ <      \  t_j  \ \ \mbox{ then } w_{j}[\ell]  \mbox{ is not significant.}
\end{array}
\end{eqnarray}

To summarize, The multiresolution support is obtained from the signal $Y$ by computing the forward transform 
coefficients, applying hard thresholding, and recording the coordinates of the retained coefficients.
 
%%%%%%%%%%%%%%%%%%%%%%%%%%%%%%%%%%%%%%%%%%%%%%%%%%%%%%
\subsection{Power Spectrum Recovery Algorithm}

Let's now turn to the adaptive estimator of the true CMB power spectrum $C[\ell]$ from its empirical 
estimate $\widehat{C}[\ell]$. As we benefit from the (asymptotic) normality of the noise in the stabilized 
samples $C^{s}[\ell]$ in \eqref{eq_logvar}, we are in position to easily construct the multiresolution 
support $M$ of $C^{s}$ as described in the previous section. Once the support $M$ of significant 
coefficients has been determined, our goal is reconstruct an estimate $\widetilde{X}$ of the true 
power spectrum, known to be sparsely represented in some dictionary $\Phi$(regularization), 
such that the significant transform coefficients of its stabilized version reproduce those of
$C^{s}$ (fidelity to data). Furthermore, as a power spectrum is a positive, a positivity constraint 
must be imposed. These requirements can be cast as seeking an estimate that solves the 
following constrained optimization problem:
\begin{equation}
\label{eq_min_supp}
\min_{X} \| { \Phi}^{T}{X}\|_1 \quad \mathrm{s.t.} \quad 
\begin{cases}  
X \geqslant 0 \\  
M  \odot \big(\Phi^{T}{\cal T}(X)\big)    =  M  \odot \big(\Phi^{T} C^{s}\big)
\end{cases},
\end{equation}
where $\odot$ stands for the Hadamard product (i.e. entry-wise multiplication) of two vectors. This problem has a global minimizer which is bounded. However, beside non-smoothness of the $l_1$-norm and the constraints, the problem is also non-convex because of the VST operator ${\cal T}$. It is therefore far from obvious to solve.

In this paper we propose the following scheme which starts with an initial guess of the power spectrum ${X}^{(0)} = 0$, and then iterates for $n=0$ to $N_{\max}-1$, 
\begin{equation}
\label{eq_iter_theo_powspec}
\begin{split}
\widetilde{{X}} &= {\cal R} \left(    {\cal T} \left( { X}^{(n)}\right)    + { \Phi} M \odot \left({ \Phi}^{T} \left(  C^{s}    -{\cal T} \left( { X}^{(n)}\right)   \right)   \right)\right) \\
{X}^{(n+1)} &= \mathcal{P}_{+}\left(  { \Phi} ~ \text{ST}_{\lambda_n}({ \Phi}^{T}\widetilde{{X}}) \right)\;,
\end{split}
\end{equation}
where $\mathcal{P}_{+}$ denotes the projection on the positive orthant and guarantees non-negativity of the spectrum estimator, $\text{ST}_{\lambda_n}(w) = \left(\text{ST}_{\lambda_n}(w[i])\right)_i$ is the soft-thresholding with threshold $\lambda_n$ that applies term-by-term the shrinkage rule
\begin{equation}
\text{ST}_{\lambda_n}(w[i]) = 
\begin{cases} 
\mathrm{sign}(w[i])(|w[i]| - \lambda_n) & \text{if} \ |w[i]| \geqslant \lambda_n~,  \\
0 & \text{otherwise} ~.
\end{cases}
\end{equation}
Here, we have chosen a decreasing threshold with the iteration number $n$, $\lambda_n = (N_{\max}-n)/(N_{\max}-1)$.
More details pertaining to this algorithm can be found in \cite{starck:book10} 

Algorithm~\ref{alg1} summarizes the main steps of the sparse denoising algorithm. A similar approach 
was proposed in \citep{starck2009,schmitt2010} for Poisson noise removal in 2D and 3D data sets.

\begin{algorithm}[!h]
\caption{TOUSI Power Spectrum Smoothing}
\label{alg1}
\begin{algorithmic}[1]
\REQUIRE $\quad$ \\
Empirical power spectrum ${\widehat{C}}$, \\
Number of iterations $N_{\max}$, \\
Threshold $\kappa$  (default value is 5).\\
\underline{\emph{\textbf{Detection}}} \\
\STATE Compute ${C^{s}}$ using \eqref{eq_logvar}.
\STATE Compute the decomposition coefficients $W$ of ${C^{s} }$ in $\Phi$, $W = \Phi^{T}  {C^{s}}$.
\STATE Compute the support $M$ from $W$ with the threshold $\kappa$, assuming standard additive white Gaussian noise. \\ 
\underline{\emph{\textbf{Estimation}}} \\
%\STATE Initialize $\alpha^{(0)}=0$.
\STATE Initialize $X^{(0)}=0$. %$\lambda_0 = \max(W)$.
\FOR{$n=0$ to $N_{\max}-1$}
%\STATE $\tilde{\alpha}= P_{+}(Q \alpha^{(n)})$.
%\STATE $\alpha^{(n+1)} = \text{ST}_{\lambda_n}[\tilde{\alpha}]$.
\STATE $\widetilde{{X}} =  {\cal R} \left(    {\cal T} \left( { X}^{(n)}\right) + { \Phi} M \odot \left({ \Phi}^{T} \left(  C^{s}    -{\cal T} \left( { X}^{(n)}\right)   \right)   \right) \right) $.
\STATE ${X}^{(n+1)} =  \mathcal{P}_{+}\left(  { \Phi} ~ \text{ST}_{\lambda_n}\left({ \Phi}^{T}\widetilde{{X}}\right) \right)$.
\STATE $\lambda_{n+1} = \frac{N_{\max} - (n+1)}{N_{\max} - 1}$ .
\ENDFOR
\STATE {\bf Return:} The estimate $\widetilde{X} = {X}^{(N_{\max})}$.
\end{algorithmic}
\end{algorithm}

%%%%%%%%%%%%%%%%%%%%%%%%%%%%%%%%%%%%%%%%%%%%%%%%%%%%%%
\subsection{Instrumental Noise}
In practice, the data are generally contaminated by an instrumental noise, and estimating the true 
CMB power spectrum $ C[\ell]$ from the empirical power spectrum $\widehat{C}[\ell]$ requires to 
remove this instrumental noise. The instrumental noise is assumed stationary and independent 
from the CMB. We will also suppose that we have access to the power spectrum of the noise, or 
we can compute the empirical power spectrum $\widehat{S}_N[\ell]$ of at least one realization, 
either from a JackKnife data map or from realistic instrumental noise simulations. The above 
algorithm can be adapted to handle this case after rewriting the optimizing problem as follows:
\begin{equation}
\label{eq_min_supp_noise}
\min_{X} \| { \Phi}^{T}{X} \|_1 \quad \mathrm{s.t.} \quad 
\begin{cases}  
X \geqslant 0 \\  
M  \odot \big(\Phi^{T}{\cal T}(X+\widehat{S}_N)\big)    =  M  \odot \big(\Phi^{T} C^{s}\big)
\end{cases}.
\end{equation}
Thus, \eqref{eq_iter_theo_powspec} becomes
\begin{eqnarray}
\label{eq_iter_theo_powspec_noise}
\begin{split}
\widetilde{{X}} &= {\cal R} \left(    {\cal T} \left( { X}^{(n)} + \widehat{S}_N\right)    + { \Phi} M \odot \left({ \Phi}^{T} \left(  C^{s}    -{\cal T} \left( { X}^{(n)}+\widehat{S}_N\right)   \right)   \right)\right) - \widehat{S}_N\\
{X}^{(n+1)} &= \mathcal{P}_{+}\left(  { \Phi} ~ \text{ST}_{\lambda_n}({ \Phi}^{T}\widetilde{{X}}) \right) ~.
\end{split}
\end{eqnarray}

Algorithm~\ref{alg1} can be modified accordingly.

\subsection{Combining Several Dictionnaries}
We have seen in Section~\ref{which_dico} that the WT and DCT dictionaries had complementary benefits. 
Indeed each dictionary is able to capture well features with shapes similar to its atoms. More generally, 
assume that we have $D$ dictionaries ${\Phi}_1,\cdots, { \Phi}_D$. Given a candidate signal $Y$, we 
can derive a support $M_d$ associated to each dictionary $\Phi_d$, for $d \in \{1,\cdots,D\}$. The optimization problem to solve now reads
\begin{equation}
\label{eq_min_supp_combi_noise}
\min_{X} \| { \Phi}^{T}{X} \|_1 \quad \mathrm{s.t.} \quad 
\begin{cases}  
X \geqslant 0 \\  
M_d  \odot \big(\Phi_d^{T}{\cal T}(X+\widehat{S}_N)\big)    =  M_d  \odot \big(\Phi_d^{T} C^{s}\big), ~ d \in \{1,\cdots,D\}
\end{cases}~.
\end{equation}
% and the solution is obtained by:
% \begin{eqnarray}
% \label{eq_iter_theo_powspec_combi_noise}
% \widetilde{{X}} &=& {\cal R} \left(    { X}^{(n)} + { \Phi_d} M_d { \Phi_d}^{T} \left( {\cal T} \left( { Y} \right)   -{\cal T} \left( { X}^{(n)} + \tilde{N} \right)   \right)   \right) \nonumber \\
% {X}^{(n+1)} &=& P_{+}\left[  { \Phi_d} \text{ST}_{\lambda_n}[{ \Phi_d}^{T}\widetilde{{X}}] \right]\;,
% \end{eqnarray}
Again, this is a challenging optimization problem. We propose to attack it by applying 
successively and alternatively \eqref{eq_iter_theo_powspec_noise} on each dictionary $\Phi_d$. 
Algorithm~\ref{alg1c} describes in detail the different steps.

\begin{algorithm}[!h]
\caption{TOUSI Power Spectrum Smoothing with $D$ dictionaries}
\label{alg1c}
\begin{algorithmic}[1]
\REQUIRE $\quad$ \\
Empirical power spectrum ${\widehat{C}}$, $D$ dictionaries ${ \Phi}_1,  ..., { \Phi}_D$, noise power spectrum $\widehat{S}_N$, \\
Number of iterations $N_{\max}$, \\
Threshold $\kappa$  (default value is 5).\\
\underline{\emph{\textbf{Detection}}} \\
\STATE Compute ${C^{s}}$ using \eqref{eq_logvar}.
\STATE For all $d$, compute the decomposition coefficients $W_d$ of ${C^{s}}$ in $\Phi_d$, $W_d = \Phi_d^{T}  {C^{s}}$.
\STATE For all $d$, compute the support $M_d$ from $W_d$ with the threshold $\kappa$, assuming standard additive white Gaussian noise. \\ 
%\STATE Compute $\widehat{S}_N$ from $N$ using Algorithm~\ref{alg1c}. \\ 
\underline{\emph{\textbf{Estimation}}} \\
%\STATE Initialize $\alpha^{(0)}=0$.
\STATE Initialize $X^{(0)}=0$, %$\lambda_0 = \max(\Phi^T {C[\ell]})$.
\FOR{$n=0$ to $N_{\max}-1$}
\STATE $Z_d = {X}^{(n)} $.
\FOR{$d=1$ to $D$}
%\STATE $\tilde{\alpha}= P_{+}(Q \alpha^{(n)})$.
%\STATE $\alpha^{(n+1)} = \text{ST}_{\lambda_n}[\tilde{\alpha}]$.
\STATE $\widetilde{{Z}} = {\cal R} \left(    {\cal T} \left( Z_d + \widehat{S}_N\right)  + { \Phi_d} M \odot \left({ \Phi_d}^{T} \left(  C^{s}    -{\cal T} \left( Z_d+\widehat{S}_N\right)   \right)   \right)\right) - \widehat{S}_N$.
\STATE $Z_{d+1} = \mathcal{P}_{+}\left(  { \Phi_d} ~ \text{ST}_{\lambda_n}({ \Phi_d}^{T}\widetilde{{Z}}) \right)$.
\ENDFOR
\STATE  ${X}^{(n+1)} = {Z}^{D+1} $.
\STATE $\lambda_{n+1} = \frac{N_{\max} - (n+1)}{N_{\max} - 1}$ .
\ENDFOR
\STATE Get the estimate $\widetilde{X} = {X}^{(N_{\max})}$.
\end{algorithmic}
\end{algorithm}

%%%%%%%%%%%%%%%%%%%%%%%%%%%%%%%%%%%%%%%%%%%%%%%%%%%%%%
\section{Application: Monte Carlo Simulations}
\label{sec:results}
We simulate $100$ maps from a theoretical CMB power spectrum that is calculated by CAMB. 
The power spectra of these maps are equivalent to $100$
realizations of the true CMB power spectrum --- This realized spectrum is
what we have access to in reality. 
Each of these $100$ simulated spectra are run through the TOUSI algorithm,
with the aim of recovering the theoretical spectrum from which these $100$ spectra were simulated 
(i.e. the one that was calculated by CAMB).

\begin{figure*}[htb]
\centering{
\includegraphics[width=\linewidth]{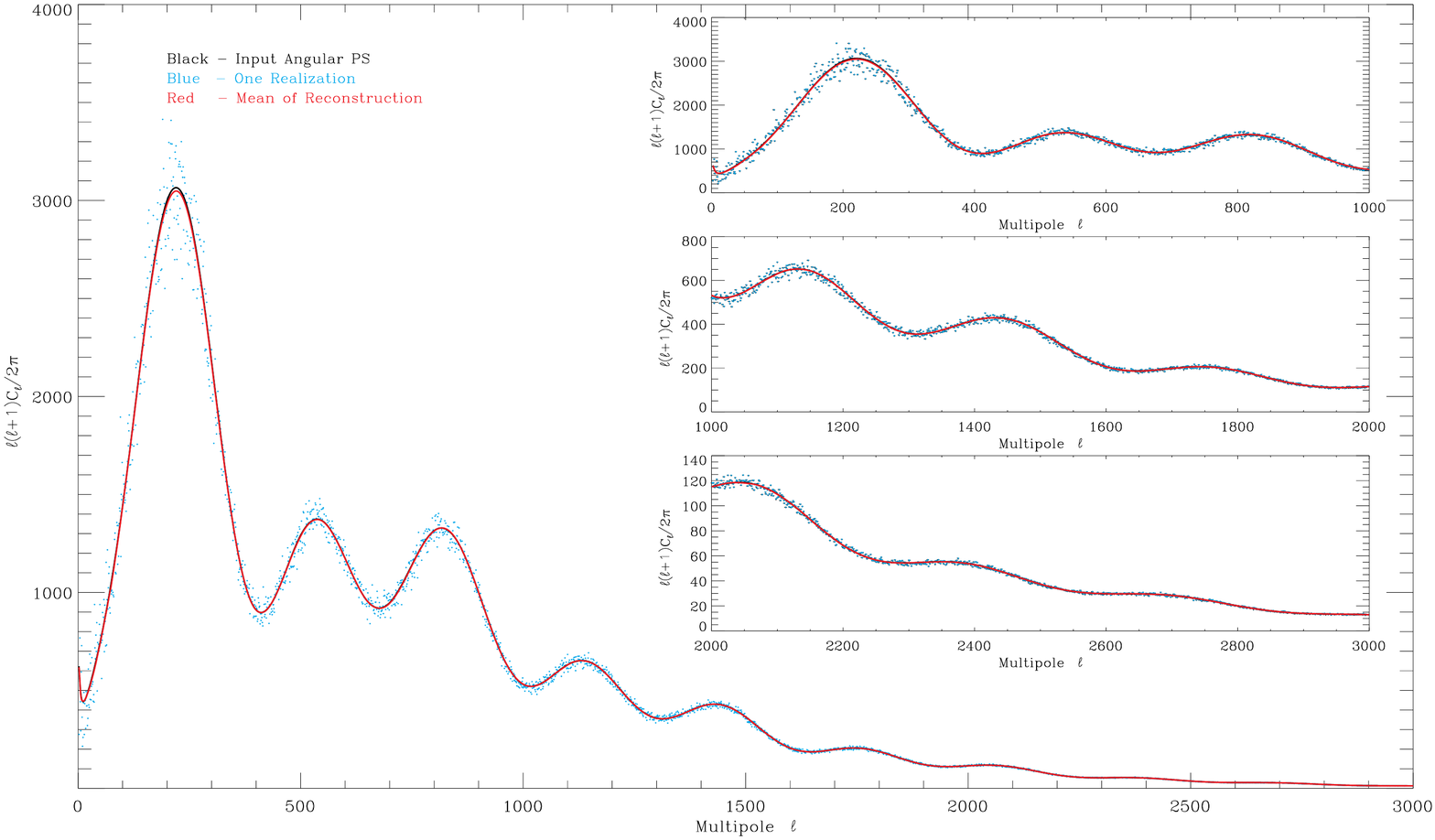}
\caption{The theoretical CMB power spectrum (black line), the empirical power spectrum of 
one realization (blue dots) and the avegared of estimated power spectra (red line) using TOUSI algorithm. 
The inner plots show a zoomed-in version.}
%\label{Fig7}
\label{Fig6}
}
\end{figure*}

\begin{figure*}[htb]
\centering{
\includegraphics[width=\linewidth]{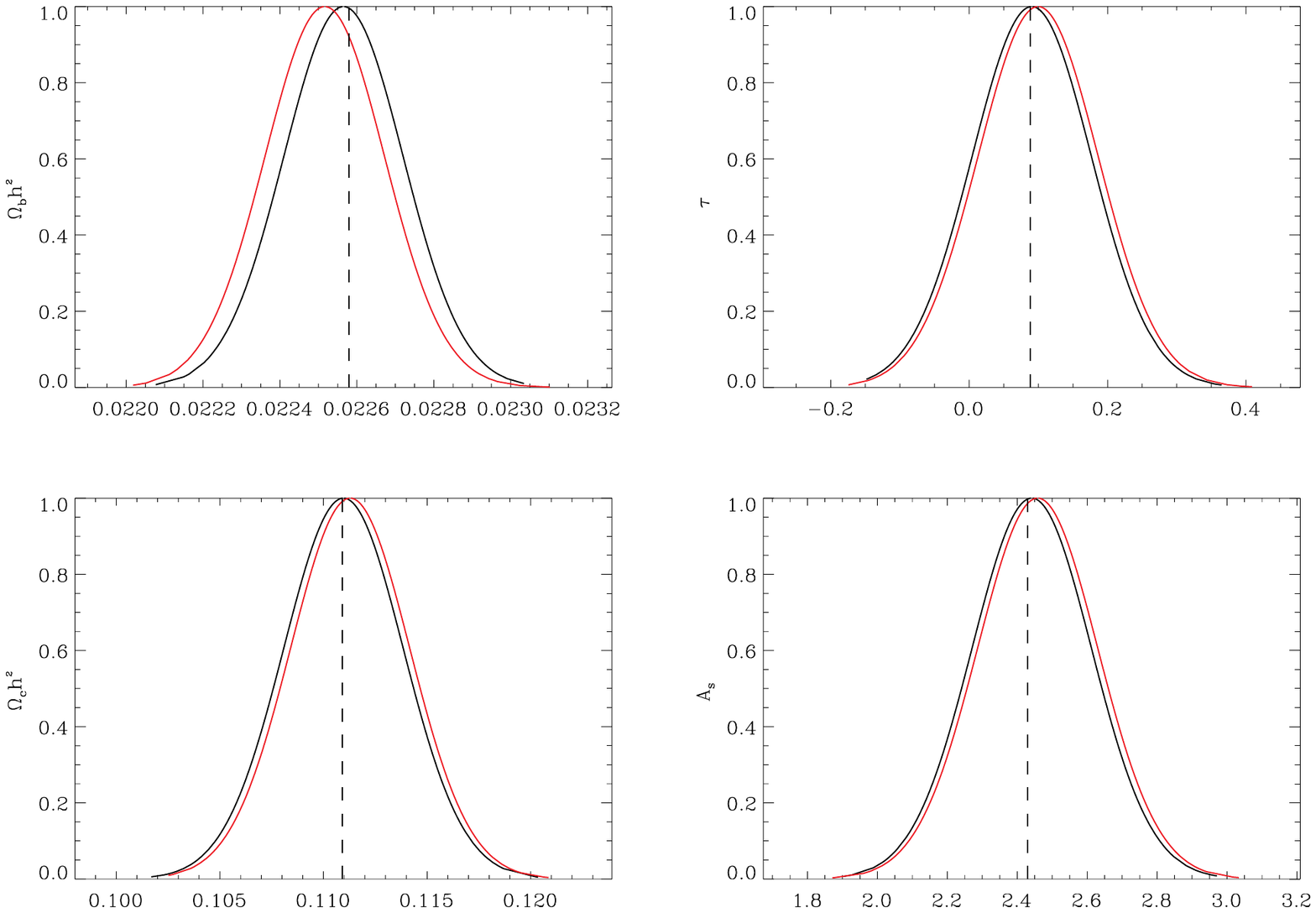}
\caption{Cosmological parameters estimated from the true CMB power spectrum
(black line) and the mean of reconstructed power spectra (red line). The dashed line is the true input parameters, i.e. the ones used to
calculate the theoretical power spectrum using CAMB.}}
%\label{Fig81D}
\label{Fig7}
\end{figure*}

Figure~\ref{Fig6} shows an example of this; the empirical power spectrum of one realization (blue dots) 
that was fed into the TOUSI algorithm, the average of the 100 estimated power spectral using TOUSI (red line) 
and the input theoretical spectra (black line). The black line is the input theoretical spectrum that was calculated by CAMB, 
which is what we are trying to recover.
%The blue dots show the simulated spectrum, which has been put through the TOUSI algorithm to give reconstructed spectrum, the red line. 
The reconstruction of the peaks and troughs of the power spectrum by this algorithm is
very impressive. This is very important as these features define the cosmological parameters.
To further check the accuracy of these reconstructed spectra, we estimate
a set of cosmological parameters from these spectra, using CosmoMC \citep{CosmoMC}.
First, a set of cosmological parameters are estimated from the $100$ simulated spectra. 
The results are shown in Figure~\ref{Fig7} as black solid lines. 
Then a `mean' reconstructed spectrum is calculated by averaging the $100$ 
reconstructed spectra. This, in principle (i.e. if the algorithm has worked), 
should be an estimation of the true input spectrum with the same characteristics 
and the same cosmological parameters. To test this, we use this `mean' reconstructed spectrum 
to simulate another $100$ maps and then $100$ spectra. These simulated spectra
are run through CosmoMC to estimate the same set of cosmological parameters. 
These are shown as red lines in Figure~\ref{Fig7}. It can be seen that 
TOUSI algorithm can reconstruct the true underlying power spectrum with great accuracy in the
cosmological parameters.

%%%%%%%%%%%%%%%%%%%%%%%%%%%%%%%%%%%%%%%%%%%%%%%%%%%%%%
\subsection{Data with Instrumental Noise}

Here we present the performance of the TOUSI algorithm in the presence of instrumental noise. 
The noise maps were simulated using a theoretical (PLANCK level) noise power spectrum. 
They were added to the CMB maps simulated previously and the power
spectra of the combined maps were estimated using equation \ref{eq:CMB_PS}. 

\begin{figure*}[htb]
\centering{
\includegraphics[width=\linewidth] {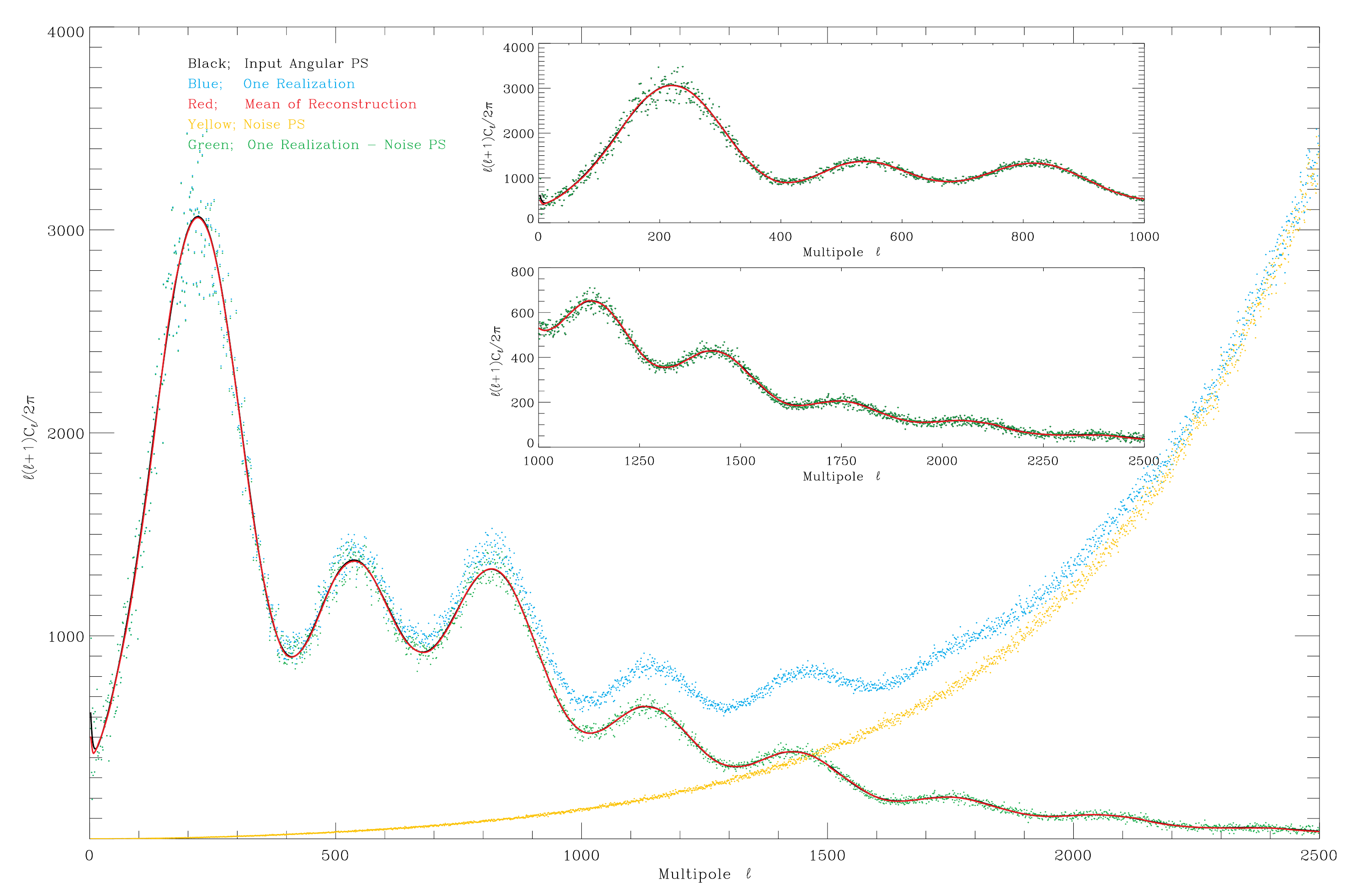}
\caption{Power spectrum estimation in the presence of instrumental noise. The blue dots show 
the empirical power spectrum of one realization having instrumental noise. 
Yellow dots show the estimated power spectrum of one of the simulated noise maps. Green dots show the 
the spectrum with the noise power spectrum removed. The black and red solid lines are the input and reconstructed 
power spectra respectively. The inner plots show a zoomed-in version.
%\label{fignoise}
\label{Fig8}
}}
\end{figure*}

Figure~\ref{Fig8} shows the reconstruction of the theoretical CMB spectrum in the presence of noise. 
The blue dots show the empirical power spectrum of one realization having instrumental noise. 
Yellow dots show the estimated power spectrum of one of the simulated noise maps. Green dots show the 
the spectrum with the noise power spectrum removed. The black and red solid lines are the input and reconstructed 
power spectra respectively. The theoretical power spectrum can be
reconstructed up to the point where the structure of the power spectrum has not been destroyed by the instrumental noise. In our case, 
having PLANCK level noise, this goes to $\ell$ up to $2500$. It can be seen that TOUSI can do a great job in reconstructing the 
input power spectrum even in the presence of instrumental noise.

%%%%%%%%%%%%%%%%%%%%%%%%%%%%%%%%%%%%%%%%%%%%%%%%%%%%%%
\subsection{Test on WMAP7 power spectrum}
\begin{figure*}[htb]
\centering{
\includegraphics[width=\linewidth]{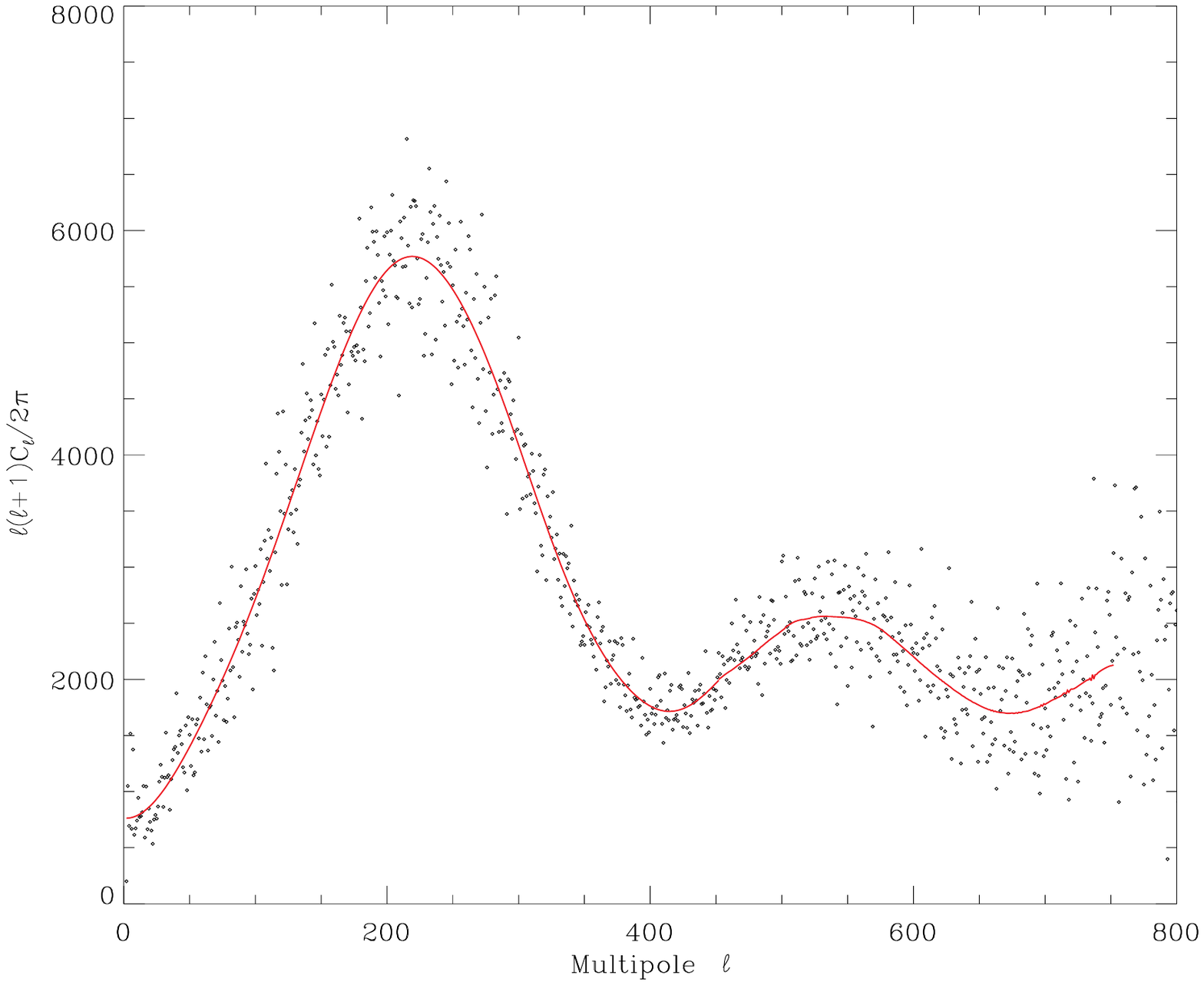}
\caption{TOUSI algorithm applied to WMAP7 power spectrum. The technique works well up to $\ell$ of $\sim800$,
i.e. before the instrumental noise becomes dominant washing out the features of the spectrum.}
%\label{Fig111}
\label{Fig9}
}
\end{figure*}

We test our algorithm on real data. Figure~\ref{Fig9} shows the application of our technique to
the WMAP7 power spectrum. The method works really well up to $\ell$ of $\sim800$. 
After this point the instrumental noise becomes so dominant that the features of the 
spectrum is washed out and cannot be recovered. 

%%%%%%%%%%%%%%%%%%%%%%%%%%%%%%%%%%%%%%%%%%%%%%%%%%%%%%
\section{Sparsity versus Averaging}
\label{sec:discussion}
A very common approach to reduce the noise on the power spectrum is the moving average filter, i.e. average values in a given window, 
\begin{equation}
\widetilde{C}^{A}[b] = \frac{1}{b(b+1)\omega_b}  \sum_{\ell=b-{\frac{\omega_b}{2}}}^{b+{\frac{\omega_b}{2}}}  \ell (\ell+1) \widehat{C}[\ell]~,
\end{equation}
and the window size $\omega_b$ is increasing with $\ell$. Here, we use window sizes of $\{1,   2,   5,  10,  20,  50,  100\}$ respectively 
for $\ell$ ranging from $\{2,  11,  31, 151, 421,  1201, 2501\}$ to $\{10, 30, 150, 420, 1200, 2500, 3200\}$, which have also been 
used in the framework of the PLANCK project in \citep{leach08}.
 
\begin{figure*}[htb]
\centering{\includegraphics[width=\linewidth]{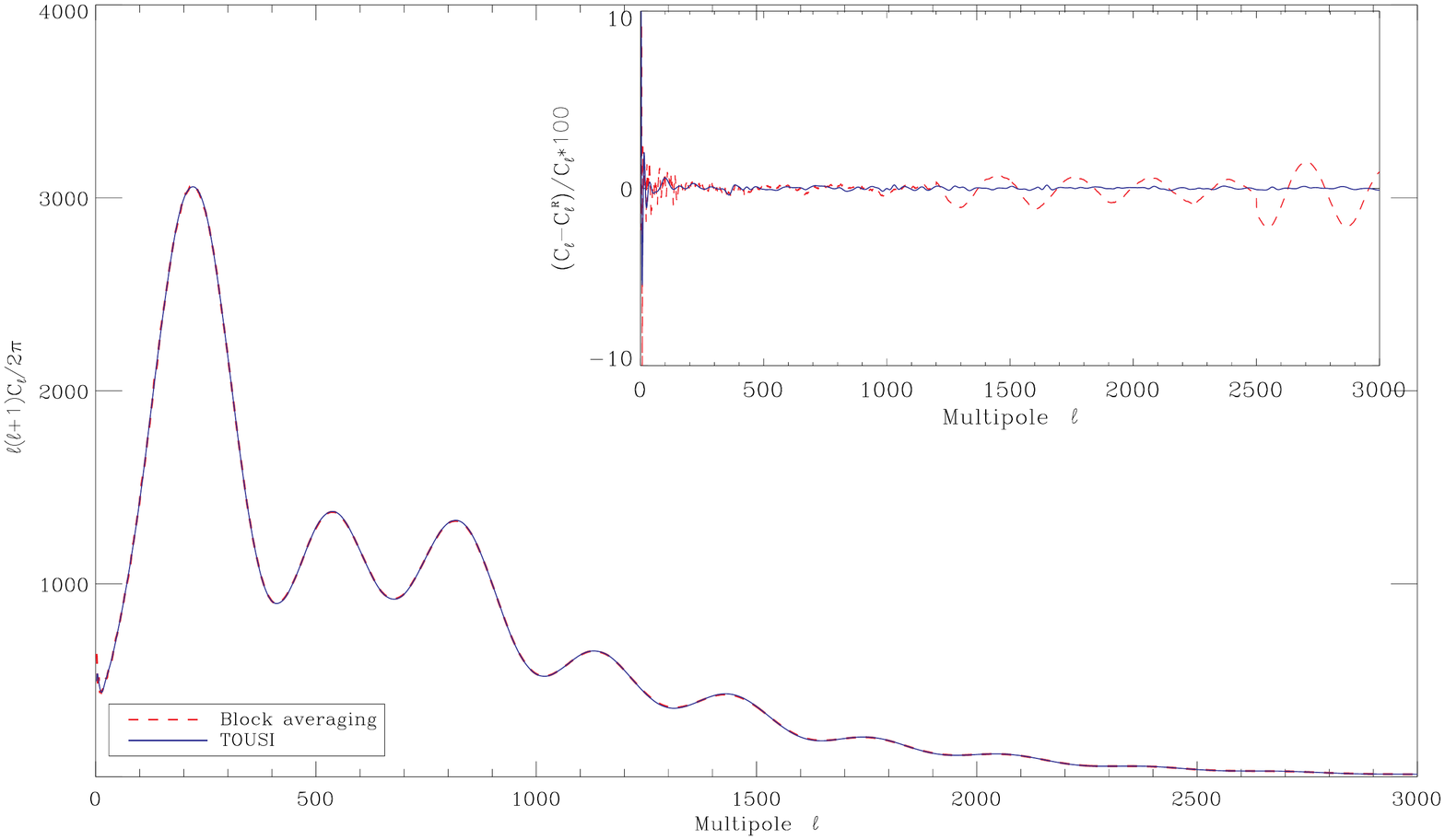}
\caption{Mean denoised spectra with Wavelets (blue solid line) and averaging (red dashed line) from 100 realizations.
The inner plot shows the normalized error for both dictionaries. }
%\label{fig_block_wt}
\label{Fig10}
}
\end{figure*}

\begin{figure*}[htb]
\centering{
\includegraphics[width=\linewidth]{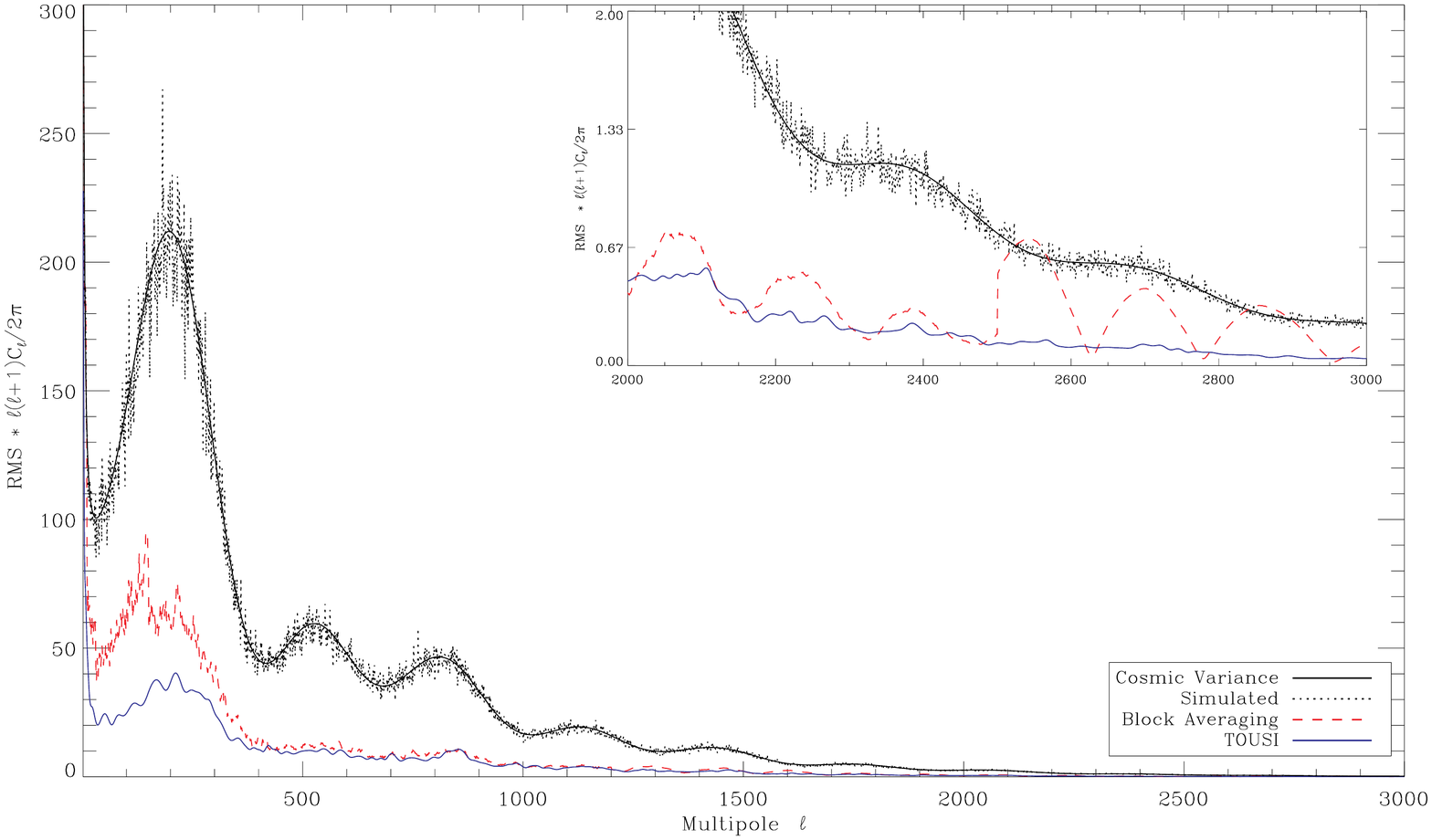}
\caption{Mean error for the 100 realizations, for the realizations (black dotted line), the averaging denoising (red dashed line) 
and the sparse wavelet filtering (blue solid line). The inner plot shows a zoom between $l=2000$ and $l=3000$.}
%\label{fig_err_block_wt}
\label{Fig11}
}
\end{figure*}

Figure~\ref{Fig10} shows a reconstruction of the power spectrum for TOUSI versus averaging. From the NMSE curve it is 
clear that our algorithm is much more efficient. Figure~\ref{Fig11} shows the average error the 100 realizations as a function of $\ell$
\begin{equation}
E[\ell] = { \frac{1}{100}  \sum_{i=1}^{100}  \parallel  C[\ell] - \widetilde{C}_i[\ell]   \parallel_2}~,
\end{equation}
where $\widetilde{C}_i$ is the estimated power spectrum from the $i$-th realization. We display the errors for the spectra 
estimated by the empirical estimator (the realization, black dotted line), the averaging estimator (red dashed line) and 
TOUSI (solid blue line). The cosmic variance is over-plotted as a solid black line. We can see that the 
expected error is highly reduced when using the sparsity-based estimator.

%%%%%%%%%%%%%%%%%%%%%%%%%%%%%%%%%%%%%%%%%%%%%%%%%%%%%%
\section{Conclusion}
\label{sec:conclusion}
Measurements of the CMB anisotropies are powerful cosmological probes.
In the currently favored cosmological model, with the nearly Gaussian-distributed
curvature perturbations, almost all the statistical information are
contained in the CMB angular power spectrum. In this
paper we have investigated the sparsity of the CMB power spectrum
in two dictionaries; DCT and WT. In both dictionaries the CMB power
spectrum can be recovered with only a few percentages of the coefficients, meaning the spectrum is very sparse. The two dictionaries have different characteristics and can accommodate reconstructing different features of the spectra; The DCT can help recover the global features of the spectrum, while WT helps recover small localized features.
The sparsity of the CMB spectrum in these two domains has helped us develop an algorithm, TOUSI, that estimates the true underlying
power spectrum from a given realized spectrum. This algorithm uses the sparsity of the CMB power spectrum in both WT and DCT domains and takes the best from both worlds to get a highly accurate estimate from a single realization of the CMB power spectrum. This could be a replacement for CAMB in cases where knowing the 
cosmological parameters is not necessary. The developed IDL code will be released with the next 
version of ISAP (Interactive Sparse astronomical data Analysis Packages) via the web site:\\ \\
{\centerline{\texttt{http://jstarck.free.fr/isap.html}}}\\

%%%%%%%%%%%%%%%%%%%%%%%%%%%%%%%%%%%%%%%%%%%%%%%%%%%%%%
\section*{Acknowledgments}
The authors would like to thank Marian Douspis, Olivier Dor\'e and Amir Hajian for useful discussions.
This work is supported by the European Research Council grant SparseAstro (ERC-228261)

% \appendix

%%%%%%%%%%%%%%%%%%%%%%%%%%%%%%%%%%%%%%%%%%%%%%%%%%%%%%
\bibliographystyle{aa}
\bibliography{JLSBibTex}

\end{document}